\documentclass[twocolumn]{aastex63}

\received{20, Dec, 2024}
\accepted{TBD}
\acceptjournal{ApJ}

\submitjournal{ApJ}

\shorttitle{Dust and gas at $z>12$}
\shortauthors{Mitsuhashi, Zavala, Bakx et al.}

\graphicspath{{./}}
\renewcommand\baselinestretch{0.92}
\usepackage{amsmath}
\tabletypesize{\footnotesize}
\usepackage{multirow}

\usepackage{ulem} 
\def\blue#1 {{\textcolor{blue}{#1}}\ }

\def\kms{\,km\,s$^{-1}$}

\def\red#1 {{\textcolor{red}{#1}}\ }

\defcitealias{2020MNRAS.495.1577I}{I20}
\begin{document}

\title{Low dust mass and high star-formation efficiency at $z>12$ from deep ALMA observations}

\correspondingauthor{Ikki Mitsuhashi}
\email{ikki0913astr@gmail.com}

\author[0000-0001-7300-9450]{Ikki Mitsuhashi}
\affiliation{Department for Astrophysical \& Planetary Science, University of Colorado, Boulder, CO 80309, USA}
\affiliation{Waseda Research Institute for Science and Engineering, Faculty of Science and Engineering, Waseda University, 3-4-1 Okubo, Shinjuku, Tokyo 169-8555, Japan}
\affiliation{National Astronomical Observatory of Japan, 2-21-1 Osawa, Mitaka, Tokyo 181-8588, Japan}

\author[0000-0002-7051-1100]{Jorge A. Zavala}
\affiliation{National Astronomical Observatory of Japan, 2-21-1 Osawa, Mitaka, Tokyo 181-8588, Japan}

\author{Tom J.L.C. Bakx}
\affiliation{Department of Space, Earth and Environment, Chalmers University of Technology, Gothenburg, Sweden}

\author{Akio K. Inoue}
\affiliation{Waseda Research Institute for Science and Engineering, Faculty of Science and Engineering, Waseda University, 3-4-1 Okubo, Shinjuku, Tokyo 169-8555, Japan}
\affiliation{Department of Physics, School of Advanced Science and Engineering, Faculty of Science and Engineering, Waseda University, 3-4-1, Okubo, Shinjuku, Tokyo 169-8555, Japan}

\author{Marco Castellano}
\affiliation{INAF - Osservatorio Astronomico di Roma, via di Frascati 33, 00078 Monte Porzio Catone, Italy}

\author{Antonello Calabro}
\affiliation{INAF - Osservatorio Astronomico di Roma, via di Frascati 33, 00078 Monte Porzio Catone, Italy}

\author{Caitlin M. Casey}
\affiliation{Department of Astronomy, The University of Texas at Austin, 2515 Speedway Boulevard Stop C1400, Austin, TX 78712, USA}

\author{Maximilien Franco}
\affiliation{Department of Astronomy, The University of Texas at Austin, 2515 Speedway Boulevard Stop C1400, Austin, TX 78712, USA}

\author{Bunyo Hatsukade}
\affiliation{National Astronomical Observatory of Japan, 2-21-1 Osawa, Mitaka, Tokyo 181-8588, Japan}
\affiliation{Graduate Institute for Advanced Studies, SOKENDAI, Osawa, Mitaka, Tokyo 181-8588, Japan}
\affiliation{Institute of Astronomy, Graduate School of Science, The University of Tokyo, 2-21-1 Osawa, Mitaka, Tokyo 181-0015, Japan}

\author{Nimish P. Hathi}
\affiliation{Space Telescope Science Institute, 3700 San Martin Drive, Baltimore, MD 21218, USA}

\author[0000-0002-2634-9169]{Ryota Ikeda}
\affiliation{National Astronomical Observatory of Japan, 2-21-1 Osawa, Mitaka, Tokyo 181-8588, Japan}
\affiliation{Department of Astronomy, School of Science, SOKENDAI (The Graduate University for Advanced Studies), 2-21-1 Osawa, Mitaka, Tokyo 181-8588, Japan}

\author[0000-0002-6610-2048]{Anton M. Koekemoer}
\affiliation{Space Telescope Science Institute, 3700 San Martin Drive, Baltimore, MD 21218, USA}

\author{Jeyhan Kartaltepe}
\affiliation{Laboratory for Multiwavelength Astrophysics, School of Physics and Astronomy, Rochester Institute of Technology, 84 Lomb Memorial Drive, Rochester, NY 14623, USA}

\author{Kirsten K. Knudsen}
\affiliation{Department of Earth and Space Sciences, Chalmers University of Technology, Onsala Space Observatory, SE-43992 Onsala, Sweden}

\author{Paola Santini}
\affiliation{INAF - Osservatorio Astronomico di Roma, via di Frascati 33, 00078 Monte Porzio Catone, Italy}

\author{Toshiki Saito}
\affiliation{Faculty of Global Interdisciplinary Science and Innovation, Shizuoka University, 836 Ohya, Suruga-ku, Shizuoka 422-8529, Japan}

\author{Elena Terlevich}
\affiliation{Instituto Nacional de Astrof\'\i sica, \'Optica y Electr\'onica,Tonantzintla, Puebla, M\'exico}
\affiliation{Institute of Astronomy, University of Cambridge, Cambridge, CB3 0HA, UK}
\affiliation{Facultad de Astronomía y Geofísica, Universidad de La Plata, La Plata, Argentina}

\author{Roberto Terlevich}
\affiliation{Instituto Nacional de Astrof\'\i sica, \'Optica y Electr\'onica,Tonantzintla, Puebla, M\'exico}
\affiliation{Institute of Astronomy, University of Cambridge, Cambridge, CB3 0HA, UK}
\affiliation{Facultad de Astronomía y Geofísica, Universidad de La Plata, La Plata, Argentina}

\begin{abstract}
We investigate the dust mass build-up and star formation efficiency of two galaxies at $z>12$, GHZ2 and GS-z14-0, by combining ALMA and JWST observations with an analytical model that assumes dust at thermal equilibrium.
We obtained $3\sigma$ constraints on dust mass of $\log M_{\rm dust}/M_{\odot}<5.0$ and $<5.3$, respectively. 
These constraints are in tension with a high dust condensation efficiency in stellar ejecta but are consistent with models with a short metal accretion timescale at $z>12$. 
Given the young stellar ages of these galaxies ($t_{\rm age}\sim10\,{\rm Myrs}$), dust grain growth via accretion may still be ineffective at this stage, though it likely works efficiently to produce significant dust in galaxies at $z\sim7$.
The star formation efficiencies, defined as the SFR divided by molecular gas mass, reach $\sim10\,{\rm Gyr}^{-1}$ in a 10\,Myr timescale, aligning with the expected redshift evolution of `starburst' galaxies with efficiencies that are $\sim0.5$--$1\,{\rm dex}$ higher than those in main-sequence galaxies.
This starburst phase seems to be common in UV-bright galaxies at $z>12$ and is likely associated with the unique conditions of the early phases of galaxy formation, such as bursty star formation and/or negligible feedback from super-Eddington accretion.
Direct observations of molecular gas tracers like [C\,{\sc ii}] will be crucial to further understanding the nature of bright galaxies at $z>12$.
\end{abstract}


\keywords{galaxies: evolution - galaxies: formation - galaxies: high-redshift}

%
%
%
%
%
%
\section{Introduction}
Dust and gas are fundamental components of the interstellar medium (ISM). 
Gas is the primary fuel for star formation and dust plays a key role in rapidly cooling this component to facilitate the collapse of gas clouds into stars. 
Cosmic dust originates from stellar phenomena, such as the ejecta of asymptotic giant branch (AGB) stars and rapidly cooling SN ejecta \citep{1980ApJ...239..193D,2001MNRAS.325..726T,2007ApJ...666..955N}.
The dust produced by stellar activity (``stellar dust'') grows in the dense ISM through metal accretion and condensation into dust grains \citep[``ISM dust'', e.g.,][]{1990ASPC...12..193D,2009ASPC..414..453D,2011EP&S...63.1027I,2014A&A...562A..76Z}.
Therefore, the dust content within galaxies is directly connected to the star formation history and the metal enrichment process.

The formation and growth of interstellar dust have been studied in relation to other galactic properties, such as the stellar mass, gas mass, and metallicity \citep{2014A&A...565A.128C,2014A&A...562A..30S,2015ApJ...806..110D,2015MNRAS.451L..70M}.
The dust-to-gas mass ratio (D/G) represents the fraction of dust relative to the ISM, serving as a powerful tool for investigating dust growth.
In the local Universe, D/G has been intensively investigated over the past decades \citep[e.g.,][]{1982A&A...107..247K,1990A&A...236..237I,2002A&A...388..439H,2010MNRAS.402.1409B,2011A&A...535A..13M} and found to be correlated with metallicity across a large dynamical range \citep{1998ApJ...496..145L,2007ApJ...663..866D,2011ApJ...737...12L,2014A&A...563A..31R}, which may reflect both chemical evolution and dust grain growth \citep{1998ApJ...501..643D,2008ApJ...672..214G,2013EP&S...65..213A,2017MNRAS.471.3152P,2020MNRAS.494.1071G}. 
Constraining D/G at high redshift is especially important, as the Universe's young age limits the contribution of several prominent dust production processes \citep[e.g.,][]{2012ApJ...760....6M,2023A&A...670A.138P,2024A&A...685A.138V,2024MNRAS.528.2407P,2024arXiv240917223F}.

Recent Atacama Large Millimeter/submillimeter Array (ALMA) studies found a substantial amount of dust at $z=6$--8 \citep{2017ApJ...837L..21L,2019ApJ...874...27T,2022MNRAS.515.3126I,2023MNRAS.523.3119W}. Several scenarios have been considered to account for this significant dust mass at this early epoch, such as high dust condensation efficiency or efficient metal accretion into ISM \citep{2019MNRAS.490.1425L,2019MNRAS.490..540L,2020MNRAS.494.1071G,2022MNRAS.512..989D,2023Natur.621..267W,2024arXiv241023959B}.
Investigation of the emergence of dust in more distant galaxies is essential for understanding the rapid increase in dust mass in the early Universe and identifying the main processes that contribute to this growth \citep[e.g.,][]{2024arXiv241023959B}.
The small age of the universe further helps explore the rapid pathways that drive dust growth.

Beyond relative dust and gas content, the total gas mass also provides valuable insights, as stars form from cold gas via various cooling mechanisms including dust thermal emission.
Specifically, the star formation efficiency (SFE), defined as the star formation rate (SFR) divided by the total (or cold) gas mass, indicates how galaxies convert gas into stars \citep[see][for review]{2020ARA&A..58..157T}. 
Observations of cold gas tracers such as CO and [C\,{\sc ii}] lines have shown that galaxies on the star-forming main sequence (MS) typically exhibit lower SFE than `starburst' galaxies \citep[e.g.,][]{1991ApJ...370..158S,2012ApJ...747L..31S,2018ApJ...853..179T}.
For instance, \citet{2020A&A...643A...5D} reported a weak redshift evolution of the gas-depletion timescale ($t_{\rm dep}\propto1/{\rm SFE}$) in $z\sim5$ UV-selected galaxies, suggesting moderate star formation activity even at high-$z$.
However, whether this weak evolution is maintained at $z>12$ is entirely unknown, but it could be a key probe to investigate the origin of the overabundance of UV-bright galaxies at $z>12$ \citep[e.g.,][]{2022ApJ...938L..15C,2023MNRAS.519.1201A,2023ApJS..265....5H,2024ApJ...969L...2F}.

In this paper, we examine the interstellar medium of two $z>12$ galaxies using ultra-deep ALMA observations of the [O\,{\sc iii}] emission and the dust continuum.
Our study focuses on two of the highest-redshift galaxies known to date, GHZ2 at $z=12.3$ \citep{2024ApJ...972..143C,2024NatAs.tmp..258Z} and GS-z14-0 at $z=14.2$ \citep{2024Natur.633..318C}.
Leveraging the gas phase and stellar metallicities as well as stellar assembly histories observed at $z>12$, which have been available thanks to the exceptional sensitivity of the {\it James Webb Space Telescope} (JWST), we aim to explore dust production, dust growth, and star formation efficiency in these early galaxies by exploiting the available deep ALMA observations.

The paper is organized as follows: Section \ref{sec:obs} provides an overview of the datasets used in this work. Section \ref{sec:analysis} describes the method of dust mass measurements. In Section \ref{sec:results}, we present our constraints and discussions from the $M_{\ast}$-$M_{\rm dust}$ relations and the gas-phase metallicity $Z$ and dust-to-gas mass relation at $z>12$, the dust production history and the star formation efficiency of the target galaxies. The conclusions are presented in Section \ref{sec:summary}. Throughout this paper, we assume a flat universe with the cosmological parameters of $\Omega_{\rm M}=0.3$, $\Omega_{\Lambda}=0.7$, $\sigma_{8}=0.8$, and $H_0=70$ \kms ${\rm Mpc}^{-1}$.

%
%
%
%
%
%
\section{Targets}\label{sec:obs}

%
%
%
%
%
%
\begin{figure*}[htbp]
\begin{center}
\includegraphics[width=18cm,bb=0 0 1000 650, trim=0 1 0 0cm]{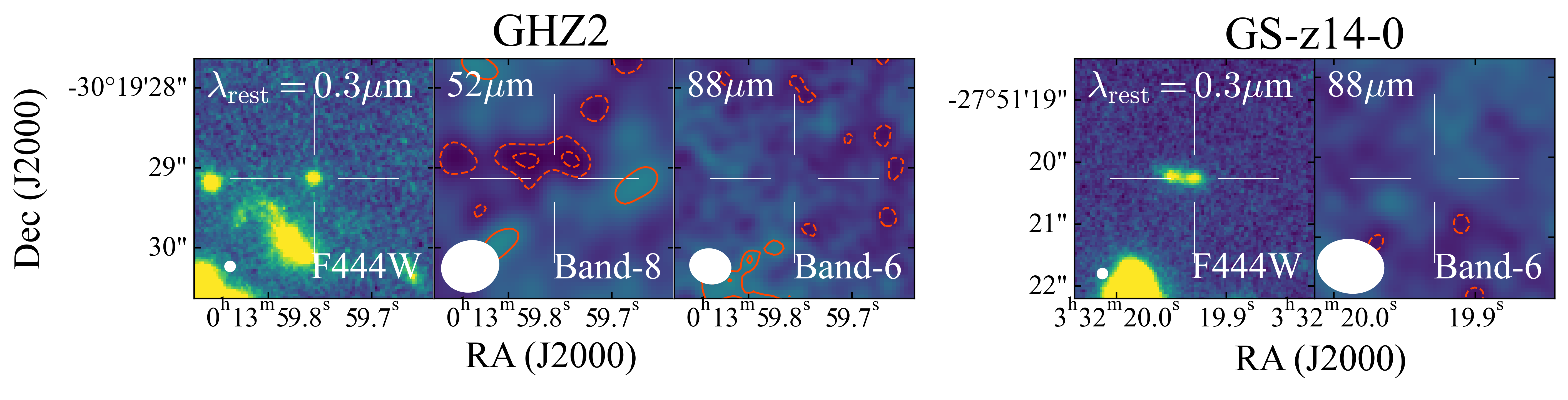}
\caption{JWST NIRCam F444W and ALMA Band-6/8 thumbnails of GHZ2 in the rest-frame $0.3\,\mu{\rm m}$, $52\,\mu{\rm m}$ and $88\,\mu{\rm m}$ continuum (left) and GS-z14-0 in the rest-frame $0.3\,\mu{\rm m}$ and $88\,\mu{\rm m}$ continuum (right). Contour levels are shown in every $2\sigma$ from $\pm3\sigma$. At the positions of the detection in $\lambda_{\rm rest}=0.3\,\mu{\rm m}$, the dust continuum remains undetected for both galaxies.}
\label{fig:thumnbnail}
\end{center}
\end{figure*}

\subsection{GHZ2}\label{subsec:ghz2}
GHZ2 was originally discovered by JWST photometry as part of the GLASS-JWST Program \citep{2022ApJ...935..110T} and identified in several studies as a robust candidate of the brightest galaxy at $z>11$ \citep{2022ApJ...938L..15C,2022ApJ...940L..14N,2023ApJS..265....5H,2023MNRAS.523.1009B}.
Its spectroscopic redshift was confirmed to be $z=12.3$ thanks to the detection of rest-frame UV/optical emission line with JWST/NIRSpec \citep{2024ApJ...972..143C} and JWST/MIRI \citep{2024NatAs.tmp..258Z}. 
More recently, \citet{2024arXiv241103593Z} reported a detection of [O\,{\sc iii}]88$\mu{\rm m}$ at $\sim5\sigma$ level and provided more accurate redshift ($z=12.3327\pm0.0005$).
Both NIRSpec and MIRI constrained its metallicity to be low \citep[$Z_{\rm gas}/Z_{\odot}\sim0.1$, see also][]{2024arXiv240312683C} and, across this paper, we will adopt the reported value of  $Z_{\rm gas}/Z_{\odot}=0.05_{-0.03}^{+0.12}$ from \citet{2024NatAs.tmp..258Z}.
The inferred stellar mass changes depending on the SED fitting code and associated assumptions, with a range of $\log M_{\ast}[M_{\odot}]=8.3$--8.9 \citep[see Appendix in][]{2024NatAs.tmp..258Z} after lens magnification correction of $\mu=1.3$\footnote{While the gravitational lens magnification factor might be up to $\mu\sim1.6$ due to the closest galaxy at $z=1.678$, we adopt $\mu=1.3$ throughout this paper following \citet{2024arXiv241103593Z}.} \citep{2023ApJ...952...84B}.
Throughout this paper, we will adopt the star-formation history derived by {\sc Bagpipes} \citep{2018MNRAS.480.4379C} with a total stellar mass of $\log M_{\ast}[M_{\odot}]=8.27_{-0.18}^{+0.23}$ owing to its non-parametric approach and good agreement with {\sc Prospector} \citep{2021ApJS..254...22J} results in \citet[][see Figure \ref{fig:productionmodel}]{2024arXiv240618352H}.
Note that despite the rich dataset it is still unclear whether an active galactic nucleus (AGN) contributes to the UV emission of GHZ2 \citep[see also,][]{2024arXiv240312683C}. And while \citet{2024arXiv241103593Z} concluded that this galaxy is likely dominated by star formation activity based on the [O\,{\sc iii}]88$\mu{\rm m}$ line width and H$\beta$ line luminosity, further observations are necessary to accurately constrain the stellar mass and potential AGN activity in GHZ2.

\subsection{GS-z14-0}\label{subsec:gsz14}
GS-z14-0 was identified in the JADES NIRCam dataset \citep{2024ApJ...970...31R} and spectroscopically confirmed via a significant detection of the Ly-$\alpha$ break with NIRSpec/PRISM \citep{2024Natur.633..318C}.
Recent ALMA observations reported successful [O\,{\sc iii}]$88\,\mu{\rm m}$ detection at $\sim6.6\sigma$ \citep{2024arXiv240920533C,2024arXiv240920549S} and confirmed to be $z=14.1796\pm0.0007$.
We will refer to the star formation history with a stellar mass of $\log M_{\ast}[M_{\odot}]=8.84_{-0.10}^{+0.09}$ (after magnification correction; $\mu=1.17$) and the stellar/gas metallicity of $Z/Z_{\odot}=0.18_{-0.01}^{+0.01}$ from the {\sc Prospector} fitting in \citet{2024arXiv240920533C}, which is consistent with the metallicity derived in \citet{2024arXiv240920549S}.

%
%
%
%
%
%
\section{Data and Analysis}\label{sec:analysis}

%
%
%
%
%
%
\begin{figure*}[htbp]
\begin{center}
\includegraphics[width=16cm,bb=0 0 1000 650, trim=0 1 0 0cm]{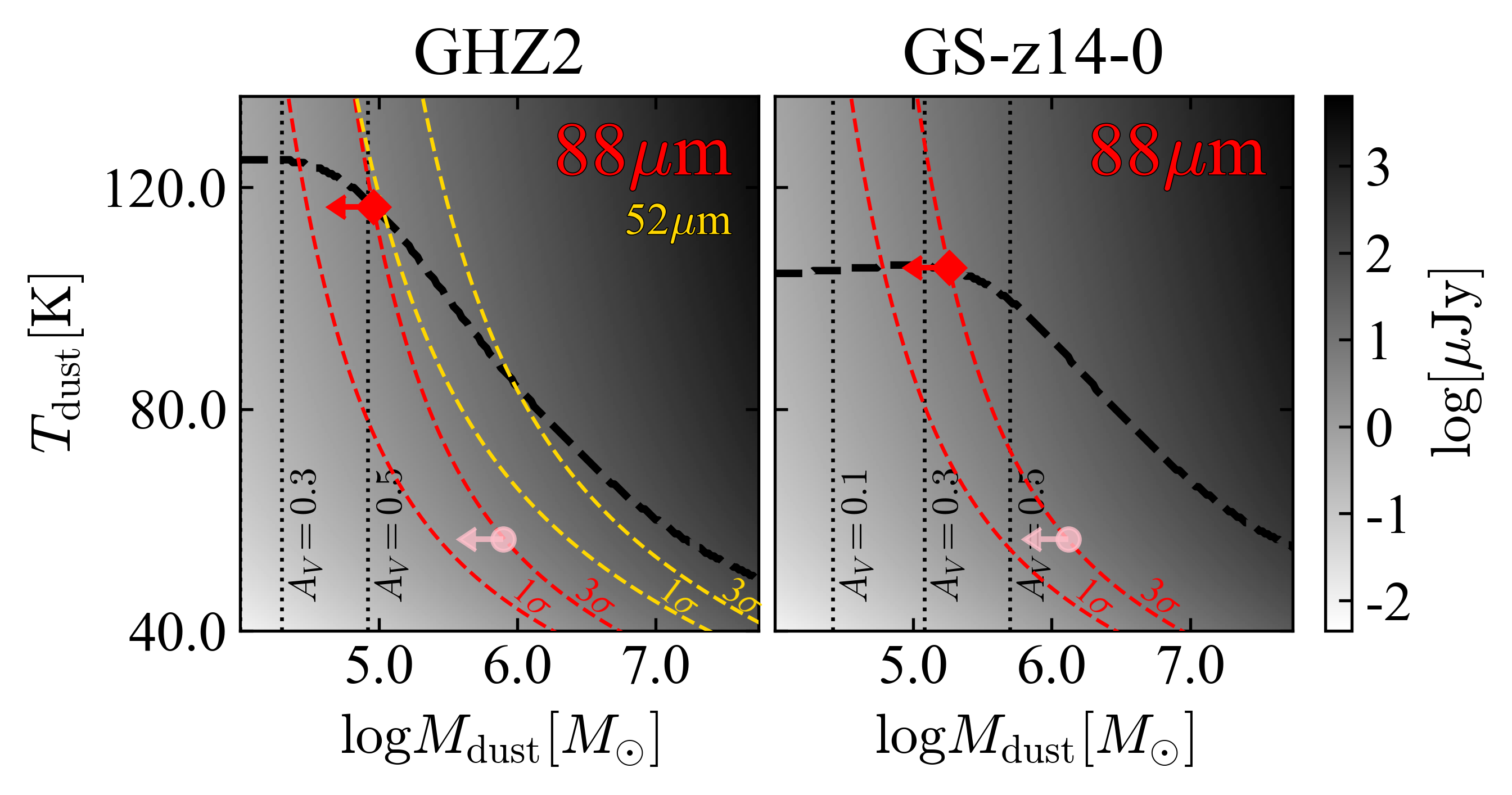}
\caption{Dust mass and temperatures, color-coded by the signal-to-noise ratio of the dust continuum that is expected at the rest-frame $88\,\mu{\rm m}$ for GHZ2 (left) and GS-z14-0 (right).  
Red solid lines indicate $1\sigma$ and $3\sigma$ upper limits at the rest-frame $88\mu{\rm m}$ by assuming a modified black body spectrum after lens magnification correction.
The yellow lines on the left panel show upper limits from the rest-frame $52\mu{\rm m}$ which provides a much less stringent constraint than those derived from the rest-frame $88\mu{\rm m}$.
The acceptable parameter sets and derived dust extinction obtained by the \citetalias{2020MNRAS.495.1577I} model with the clumpy dust geometry are overlayed in black dashed lines and dotted lines, respectively.
Our constraints based on \citetalias{2020MNRAS.495.1577I} model are shown in red diamonds in the cross-section of $3\sigma$ red solid lines and black dashed lines.
We also illustrated the $3\sigma$ upper limit under the assumption of $T_{\rm dust}=50\,{\rm K}$ in the pink circles to just demonstrate how \citetalias{2020MNRAS.495.1577I} model gives tight constraints. Note also that these dust mass constraints do not agree with the observed $A_V$ values (contrary to the values inferred using the \citetalias{2020MNRAS.495.1577I} model).}
\label{fig:params}
\end{center}
\end{figure*}

\subsection{ALMA data}\label{subsec:alma_data}
ALMA observations of GHZ2 were obtained in both Band-6 and Band-8 by two DDT programs (2021.A.00020.S; PI: Bakx and 2023.A.00017.S; PI: Zavala), where the spectral setups were designed to cover the [O\,{\sc iii}]$88\,\mu{\rm m}$ and the [O\,{\sc iii}]$52\,\mu{\rm m}$ lines, respectively.
GS-z14-0 has also been observed by ALMA as part of a DDT program (2023.A.00037.S; PI: Schouws) targeting [O\,{\sc iii}]$88\,\mu{\rm m}$ emission line in Band-6.
The calibration of the ALMA data is described in detail in \citet[][see also, \citealt{2023MNRAS.519.5076B,2024arXiv240920533C}]{2024arXiv241103593Z}, here we briefly summarize it.
The data were calibrated following standard ALMA pipeline workflow using CASA version 6.5.1 \citep{2022PASP..134k4501C}. 
After removing channels $\pm300\,{\rm km}\,{\rm s}^{-1}$ around the detected [O\,{\sc iii}]$88\,\mu{\rm m}$ and expected [O\,{\sc iii}]$52\,\mu{\rm m}$ line frequencies, we created continuum images with a natural weighting to maximize the signal-to-noise ratio.
We simply used a dirty image without a primary beam correction because the S/Ns of our target lines are not expected to be high enough to require {\sc clean}ing procedure and we require uniform noise distribution within a field of view.
The achieved r.m.s levels for GHZ2 and GS-z14-0 data are $3.4\,\mu{\rm Jy}\,{\rm beam}^{-1}$ and $5.0\,\mu{\rm Jy}\,{\rm beam}^{-1}$ with beamsizes of $0\farcs51\times0\farcs43$ and $0\farcs82\times0\farcs66$ in Band-6 before lens magnification correction, respectively.
For GHZ2, the Band-8 observations have an r.m.s depth of $54.6\,\mu{\rm Jy}\,{\rm beam}^{-1}$ with a beamsize of $0\farcs71\times0\farcs63$, which is much shallower than Band-6 depth and less constraining on $M_{\rm dust}$ even when taking $T_{\rm dust}$ variation into account.

Figure \ref{fig:thumnbnail} shows thumbnails in the rest-frame $0.3\mu{\rm m}$ and $88\mu{\rm m}$ (and $52\mu{\rm m}$ for GHZ2) taken by JWST and ALMA.
None of the dust continua were detected even with $\gtrsim7$ hours on source integrations.

\subsection{$M_{\rm dust}$ measurement}\label{subsec:dustmass}
We constrain the dust mass of these galaxies by applying an analytical model from \citet[][hereafter \citetalias{2020MNRAS.495.1577I}]{2020MNRAS.495.1577I} to the deep ALMA observations.
\citetalias{2020MNRAS.495.1577I} proposed an algorithm to determine the dust mass ($M_{\rm dust}$) and temperatures ($T_{\rm dust}$) under the assumption of radiative equilibrium.
In their model, the dust temperature relates to the input radiation source from young stars and cosmic microwave background (CMB) that are absorbed by dust.
Then, by incorporating dust geometry, we can calculate the escape probability of UV photons from interstellar media, which results in observable UV luminosities.
Among the three available models for the dust geometry provided in \citetalias{2020MNRAS.495.1577I} \citep[see also,][]{2018ApJ...854...36I}, we specifically employed the ``clumpy'' geometry as it is considered a reliable approximation for high-$z$ galaxies \citep[e.g.,][]{2023MNRAS.521.2962F,2023MNRAS.520L..16K}.
Indeed, this `clumpy' geometry provided looser constraints on $M_{\rm dust}$ than other geometries in \citetalias[][i.e., `spherical' or `shell']{2020MNRAS.495.1577I} under the same assumptions, and therefore our analysis is considered to be conservative.
We assumed the dust to be optically thin with an emissivity index of $\beta_{\rm dust}=2.0$ \citep{2012MNRAS.425.3094C,2021ApJ...919...30D}, a mass absorption coefficient of $\kappa_{\rm UV} = 5.0\times10^4\,{\rm cm}^2\,{\rm g}^{-1}$, dust emissivity of $\kappa_0 = 30\,{\rm cm}^2\,{\rm g}^{-1}$ at $100\,\mu{\rm m}$, and a clumpiness parameter of $\log\xi_{\rm cl}=-1.0$ following \citetalias{2020MNRAS.495.1577I} and \citet{2023MNRAS.521.2962F}.
The only input from the observations, beyond the continuum flux density, is the spatial extent of dust.
Here, we adopt the spatial extent of the stars traced by the NIRCam photometry as those of the dust emitting regions ($r_{e, {\rm dust}}=r_{e, {\rm UV}}$, see Table \ref{tab:tab1}). 
Note that in the case of GHZ2, if we use the effective radius of $r_{e,{\rm UV}}=39\pm10\,{\rm pc}$ from \citet{2023ApJ...951...72O} instead of $105\pm9\,{\rm pc}$, the $M_{\rm dust}$ and $M_{\rm gas}$ constraints become tighter than the fiducial values owing to higher $T_{\rm dust}$ and smaller $M_{\rm dyn}$, respectively. 
Hence, the results presented here are considered to be conservative. 

In Figure \ref{fig:params}, we show the resulting $T_{\rm dust}$ and $M_{\rm dust}$ parameter sets acceptable in our calculations based on the \citetalias{2020MNRAS.495.1577I} model, with observational constraints corrected for the gravitational magnification  ($\mu=1.3$ and $\mu=1.17$ for GHZ2 and GS-z14-0, respectively).
The upper limits on $M_{\rm dust}$ are shown at the intersection between the analytical predictions (thick black dashed line) and the observational limits (red dashed line).
Our calculations support very high $T_{\rm dust}$ ($>90\,{\rm K}$) due to the compact distribution and strong UV radiation.
A similar high $T_{\rm dust}$ was also inferred for a $z\sim10$ galaxy based on recent ALMA observations  \citep{2023ApJ...950...61Y}.
Metal-poor ISM conditions in the target galaxies may also contribute to these elevated dust temperatures
\citep{2019ApJ...874...27T,2020MNRAS.493.4294B,2022MNRAS.513.3122S,2024ApJ...971..161M}.
As a consequence of the high dust temperature conditions, our fitting set a tight $3\sigma$ upper limits on the dust masses of $\log M_{\rm dust}/M_{\odot}<5.0$ and $<5.3$ for GHZ2 and GS-z14-0, respectively. These upper limits agree well with the expected dust extinction ($A_V\lesssim0.5\,{\rm mag}$) under the `clumpy' dust geometry in each $M_{\rm dust}$ calculated through the escaped/intrinsic UV radiation. In addition, the mass constraints are in good agreement with those obtained by \citet{2024arXiv241019042F},  where they adopt the SED-based extinction as a proxy of the dust mass.

\begin{deluxetable}{cccc}
\tablecaption{Physical parameters of two $z>12$ galaxies \label{tab:tab1}}
\tablewidth{0pt}
\tablehead{
\colhead{ } & 
\colhead{GHZ2} &
\colhead{GS-z14-0} & 
\colhead{ref}
}
\startdata
$M_{\rm UV}[{\rm mag}]$ & $-20.53\pm0.01$ & $-20.81\pm0.16$ & (1,2)\\
$\mu$ & $1.3^{\small \ddag}$ & 1.17 & (1,2)\\
$r_{e, {\rm F200W}}[{\rm pc}]$ & $105\pm9$ & $260\pm20$ & (3,2)\\
$\log M_{\ast}[{\rm M}_{\odot}]$ & $8.27_{-0.18}^{+0.23}$ & $8.84_{-0.10}^{+0.09}$ & (4,2)\\
$\log M_{\rm dyn}[{\rm M}_{\odot}]$ & $8.9\pm0.2$ & $9.0\pm0.2$ & (5,2)\\
${\rm SFR}_{\rm 10Myr}[{\rm M}_{\odot}\,{\rm yr}^{-1}]$ & $8.7^{+1.4}_{-1.9}$ & $9.9^{+2.3}_{-2.7}$ & (4,6)\\
${\rm SFR}_{\rm 100Myr}[{\rm M}_{\odot}\,{\rm yr}^{-1}]$ & $1.4^{+2.0}_{-0.3}$ & $5.6^{+1.2}_{-1.3}$ & (4,6)\\
$S_{88\mu{\rm m}}[\mu{\rm Jy}\,{\rm beam}^{-1}]^{\small \dagger}$ & $<10.3$ (3.4) & $<14.9$ (5.0) & (7)\\
$S_{52\mu{\rm m}}[\mu{\rm Jy}\,{\rm beam}^{-1}]^{\small \dagger}$ & $<163.9$ (54.6) & - & (7)\\
$\log M_{\rm dust}[{\rm M}_{\odot}]^{\small \P}$ & $<5.0$ & $<5.3$ & (7)\\
\enddata
\tablecomments{Every parameter except for the ALMA observational constraints is corrected for magnification shown in the second row.
\vspace{-6pt}\tablenotetext{\small \ddag}{As described in \citet{2024NatAs.tmp..258Z}, the magnification for GHZ2 might be $\sim1.6$ if we take into account the closest galaxy at $z=1.682$.}
\vspace{-6pt}\tablenotetext{\small \dagger}{$3\sigma$ upper limits with $1\sigma$ uncertainties, without correcting the magnification}
\vspace{-6pt}\tablenotetext{\small \P}{$3\sigma$ upper limits on the dust masses assuming thermal equilibrium model in \citetalias{2020MNRAS.495.1577I}.}
\vspace{-6pt}\tablenotetext{}{1. \citet{2024ApJ...972..143C}, 2. \citet{2024Natur.633..318C}, 3. \citet{2022ApJ...938L..17Y}, 4. \citet{2024NatAs.tmp..258Z}, 5. \citet{2024arXiv241103593Z} , 6. \citet{2024arXiv240920533C}, 7. This work}
}
\end{deluxetable}

%
%
%
%
%
%
\section{Results and Discussion}\label{sec:results}

\subsection{Dust production and evolution at cosmic dawn}\label{subsec:dustprod}

%
%
%
%
%
%
\begin{figure*}[htbp]
\begin{center}
\includegraphics[width=18.5cm,bb=0 0 1000 650, trim=0 1 0 0cm]{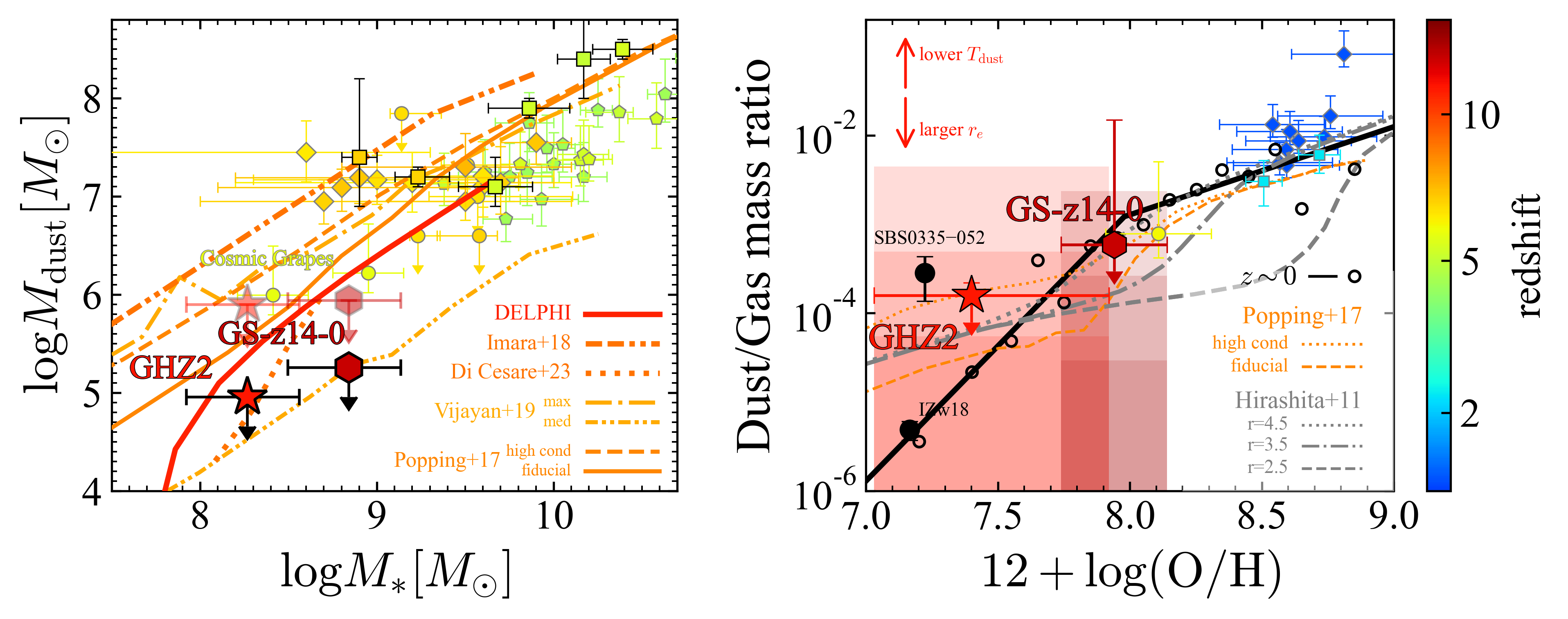}
\caption{(left) Dust mass as a function of the stellar mass. 
Our constraints for the galaxies at $z>12$ along with the previous constraints at $z=4$--8 \citep{2021MNRAS.508L..58B,2022MNRAS.513.3122S,2022MNRAS.517.5930S,2022MNRAS.515.1751W,2023MNRAS.521.2962F,2024A&A...685A.138V} and (semi-)analytical models \citep{2017MNRAS.471.3152P,2018ApJ...854...36I,2019MNRAS.489.4072V,2023MNRAS.519.4632D} including DELPHI model \citep{2023MNRAS.526.2196M} are shown in colors corresponding to their redshift. 
For \citet{2017MNRAS.471.3152P} and \citet{2019MNRAS.489.4072V} models, predictions with different dust condensation efficiencies are shown in different line styles.
The markers surrounded by black lines are constraints using the \citetalias{2020MNRAS.495.1577I} model, whereas those surrounded by gray lines show results by assuming a modified black body radiation with a typical dust temperature (i.e. $T_{\rm dust}=50\,{\rm K}$).
(right) Dust-to-gas mass ratio against the gas-phase metallicity. 
Our fiducial constraints from $M_{\rm gas}=M_{\rm dyn}-M_{\ast}$ are shown in a red star and brown hexagon, and those from $M_{\rm gas}=0.1$, 0.5, and $0.9\,(M_{\rm gas}+M_{\ast}$) are illustrated in red and brown shaded regions with different levels of opacity.
Observational constraints at $z=2$--6 \citep{2012ApJ...760....6M,2023A&A...670A.138P,2024A&A...685A.138V} and a model \citep{2017MNRAS.471.3152P} are shown in colors depending on their redshift. The black solid line and circles indicated median relation and observations at $z=0$ \citep{2014A&A...563A..31R}.
The gray lines show models with the power law in the dust size distribution of $r=4.5$ (dotted), 3.5 (dash-dotted), and 2.5 (dashed), respectively, from \citet{2011MNRAS.416.1340H}.
}
\label{fig:metalduststar}
\end{center}
\end{figure*}

In the left panel of Figure \ref{fig:metalduststar}, we compare our constraints on the dust mass and stellar mass relation with previous constraints for galaxies at $z=4$--7 \citep{2021MNRAS.508L..58B,2022MNRAS.513.3122S,2022MNRAS.517.5930S,2022MNRAS.515.1751W,2024A&A...685A.138V}.
Our results indicate a lower dust content in galaxies at $z>12$ compared to those at $z=4$--7, a trend that persists when compared with samples from \citet{2023MNRAS.521.2962F}, which were derived using a similar methodology.
This conclusion holds even when assuming $T_{\rm dust}=50\,{\rm K}$ \citep[e.g.,][]{2022MNRAS.513.3122S,2022MNRAS.517.5930S,2024ApJ...971..161M}, implying a dust mass limit about $\sim1\,{\rm dex}$ larger than our fiducial value, as shown in Figure \ref{fig:metalduststar}.
A recent study in \citet{2024arXiv241023959B} suggested a potential decrease of the dust-to-stellar mass ratio at $z>10$, which is in line with our conclusion.

The dust-to-stellar mass ratio ($M_{\rm dust}/M_{\ast}$) is likely influenced by grain growth within the ISM, which is driven by the condensation of metals in dense gas and is thus inversely proportional to gas-phase metallicity \citep[e.g.,][]{2014MNRAS.445.3039D,2024arXiv241023959B}. 
Given the positive correlation between stellar mass and gas-phase metallicity (i.e., the mass-metallicity relation), grain growth becomes effective around stellar masses of $\log M_{\ast}/M_{\odot}\gtrsim8.5$ \citep{2015MNRAS.451L..70M}. 
Models that assume very efficient grain growth due to enhanced gas condensation do not align closely with the non-detection of dust continua at $z>12$ \citep{2017MNRAS.471.3152P,2019MNRAS.489.4072V}.

In the left panel of Figure \ref{fig:metalduststar}, we also plot several (semi-)analytical models of dust production at $z>6$ \citep{2017MNRAS.471.3152P,2018ApJ...854...36I,2019MNRAS.489.4072V,2023MNRAS.526.2196M,2023MNRAS.519.4632D}, which predict dust content evolution due to production by stars, destruction by SNe, and grain growth in clouds. 
These models reproduce observed trends and scatter well at $z=4$--7, though most seem to overestimate our measurements at $z>12$, potentially because their models focus on galaxies at $z\sim7$--9.

Another possible reason for low dust content at $z>12$ is that dust is expelled or lifted off by the radiation-driven outflows \citep{2024A&A...684A.207F,2024arXiv240917223F}.
In their model, the super-Eddington star formation occurs with significant radiation pressure.
In this case, the total dust mass might be underestimated because such expelled dust is no longer in equilibrium.
Indeed, GHZ2 is derived to have specific SFR of $\log{\rm sSFR}\sim1.67\,{\rm Gyr}$, which exceeds a critical sSFR ($\log\,{\rm sSFR}\sim1.4\,{\rm Gyr}$) presented in \citet{2024A&A...684A.207F}.
While GS-z14-0 is expected to have $\log\,{\rm sSFR}\sim1.3\,{\rm Gyr}$ within recent $\sim100\,{\rm Myr}$ based on its SFH, such a radiation-driven outflow may play an important role for non-detections of the dust continuum.

%
%
%
%
%
%
\begin{figure*}[htbp]
\begin{center}
\includegraphics[width=15cm,bb=0 0 1000 650, trim=0 1 0 0cm]{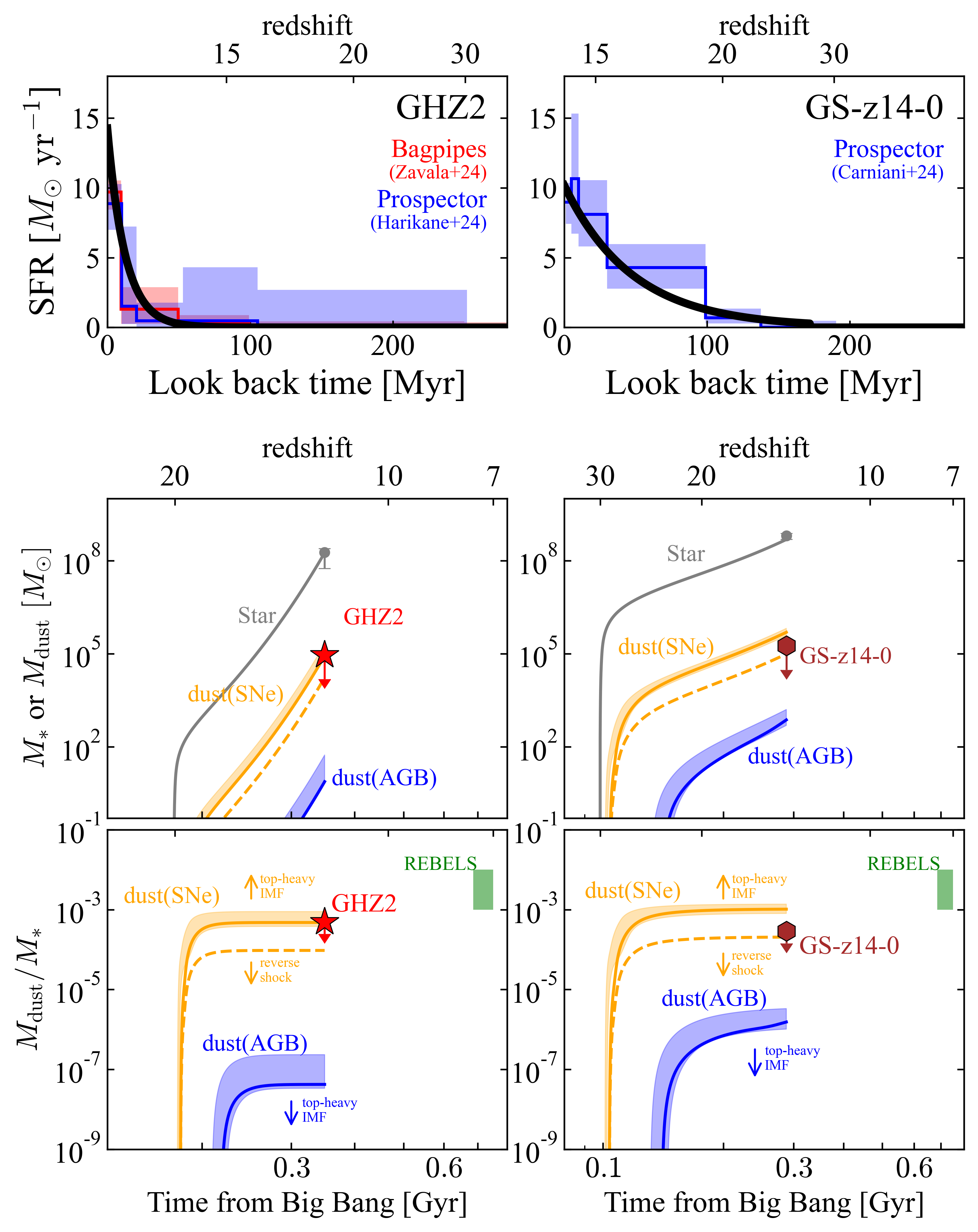}
\caption{Dust and star formation history derived from non-parametric SFH and SN/AGB dust yield models for GHZs (left) and GS-z14-0 (right). (top) Smoothed exponentially rising star formation history (black) derived from {\sc Bagpipes} \citep[red,][]{2024NatAs.tmp..258Z} and {\sc Prospector} \citep[blue,][]{2024arXiv240920533C,2024arXiv240618352H}. 
(middle, bottom) Our constraints on the dust mass with the previous results from the REBELS survey \citep{2022MNRAS.516..975T,2022MNRAS.512..989D,2022MNRAS.512...58F}. The models of the dust production history from SFH are based on \citet{2009MNRAS.397.1661V} with the SN dust yield from \citet{2007MNRAS.378..973B} and \citet{2008A&A...479..453Z}, and the AGB dust yield from \citet{2006A&A...447..553F} and ATON code \citep{2019MNRAS.486.4738D}.
For \citet{2008A&A...479..453Z} and ATON models, we show ranges with $10^{-4}$--$0.1\,Z_{\odot}$ with orange and blue shaded regions.
The effect of the initial mass function (IMF) and reverse shock is shown in two arrows.
The SN yield by \citet{2007MNRAS.378..973B} model with $20\%$ dust survival fraction is shown in the orange dashed line.} 
\label{fig:productionmodel}
\end{center}
\end{figure*}

In the right panel of Figure \ref{fig:metalduststar}, we explore the relationship between gas-phase metallicity ($Z_{\rm gas}$) and D/G.
The constraints on D/G at low metallicity values are critical to disentangle dust growth scenarios.
Here, we computed the gas mass by subtracting the magnification-corrected stellar mass from the dynamical mass ($M_{\rm gas}=M_{\rm dyn}-M_{\ast}$) assuming negligible dark matter contribution within galaxy centers ($\lesssim r_e$). 
For GHZ2 and GS-z14-0, the resulting gas masses in our fiducial caluclation are $M_{\rm gas}=5.8^{+3.7}_{-3.7}\times10^{8}\,M_{\odot}$ and $M_{\rm gas}=2.6^{+5.2}_{-2.6}\times10^{8}\,M_{\odot}$, respectively.

It is worth noting that, the mass of the ionized gas component ($M_{\rm ion}$) has an insignificant contribution.
Based on $L_{{\rm H}\beta}-M_{\rm ion}$ relation presented in \citet{2014MNRAS.442.3565C}, GHZ2 has $M_{\rm ion}=2.5\times10^6-9.8\times10^7\,M_{\odot}$ from the  ${\rm H}\beta$ flux of $F_{{\rm H}\beta}=0.9\pm0.2\times10^{-18}\,{\rm erg}\,{\rm s}^{-1}\,{\rm cm}^{-2}$ and electron density of $n_e=100-4000\,{\rm cm}^{-3}$ \citep{2024NatAs.tmp..258Z}. GS-z14-0 is expected to have $M_{\rm ion}=3.0\times10^5-1.2\times10^7\,M_{\odot}$ from $F_{{\rm H}\beta}=7.9_{-0.18}^{+0.17}\times10^{-19}\,{\rm erg}\,{\rm s}^{-1}\,{\rm cm}^{-2}$ \citep{2024arXiv240518462H}, assuming of the same range of electron density as GHZ2. 
We adopt $M_{\rm ion}=1.2\times10^7\,M_{\odot}$ as the lower limit of $M_{\rm gas}$ for GS-z14-0, as $M_{\rm gas}=M_{\rm dyn}-M_{\ast}=0$ is unphysical.

We obtained constraints of ${\rm D/G}\lesssim10^{-4}$ at $Z\sim0.1\,Z_{\odot}$ and ${\rm D/G}\lesssim10^{-3}$ at $Z\sim0.2\,Z_{\odot}$.
As described in \citet{2017MNRAS.471.3152P}, the D/G ratio as a function of metallicity is determined by the balance of the condensation efficiency ($f_{\rm cond}$) and accretion timescales ($\tau_{\rm acc}$) since both condensation and accretion promote the increase of the D/G.
Among their calculations, models with high $f_{\rm cond}$ do not align well even in the extreme case with 100\% condensation and no accretion (`no-acc' model, although their calculations are up to $z\sim9$).
At the metallicity range of $Z<0.5\,Z_{\odot}$, their fiducial model follows the fixed accretion timescale $\tau_{\rm acc}=100\,{\rm Myr}$ (`fix-tau' model), which is consistent with our constraints.
As a consequence, the inferred upper limits favor a moderate $f_{\rm cond}$ rather than the high-efficiency model in \citet{2017MNRAS.471.3152P}.

In the right panel of Figure \ref{fig:metalduststar}, we compare our target galaxies with local dwarf galaxies, SBS0335-052 and I\,Zw\,18 \citep{2016MNRAS.457.1842S}.
These two galaxies show a dramatic difference in D/G at similar gas-phase metallicity.
\citet{2016MNRAS.457.1842S} found that ISM density causes this variety because of the efficient grain growth in high-density ISM.
Further investigation on the ISM density, such as the molecular gas density (or electron density as a proxy), might be helpful to confirm the potential inefficient dust grain growth of galaxies at $z>12$.

As \citet{2019MNRAS.490..540L} suggested, both high $f_{\rm cond}$ and short $\tau_{\rm acc}$ can account for large dust amounts in massive galaxies at $z\sim7$ \citep{2019PASJ...71...71H,2019ApJ...874...27T,2020MNRAS.493.4294B}. 
In their results, the high $f_{\rm cond}$ model predicts large dust mass at the very early stage of the galaxy formation ($\lesssim10\,{\rm Myr}$) while the short $\tau_{\rm acc}$ model imply a rapid increase of dust mass at the age of $\sim100\,{\rm Myr}$.
If our target galaxies at $z>12$ are progenitors of these $z\sim7$ massive galaxies, our results support the short $\tau_{\rm acc}$ model with $\tau_{\rm acc}\sim5$--$100\,{\rm Myr}$, while the high $f_{\rm cond}$ model is unsupported. 
From another perspective, the models by \citet{2011MNRAS.416.1340H} reveal that dust size distribution and $\tau_{\rm acc}$ significantly affect D/G at metallicities of $Z\sim0.1$--$2\,Z_{\odot}$ \citep[see also,][]{2013EP&S...65..213A}. 
Our results at $Z\sim0.2\,Z_{\odot}$, while having a large uncertainty, suggest that a moderately large-size dust distribution with $\tau_{\rm acc}\sim10\,{\rm Myr}$ can consistently explain the low D/G observed at low metallicities, as shown in Figure \ref{fig:metalduststar}.

Here, we emphasize the importance of the direct observations of molecular gas tracers.
In the right panel of Figure \ref{fig:metalduststar}, we also illustrate the acceptable D/Gs based on $M_{\rm gas}$ assuming gas mass fractions of $f_{\rm gas}=M_{\rm gas}/(M_{\rm gas}+M_{\ast})=0.1$, 0.5, and 0.9 instead of those from $M_{\rm dyn}-M_{\ast}$.
Nevertheless, it is impossible to disentangle any models with this broad $f_{\rm gas}$ range of 0.1--0.9.
Further ALMA observations targetting [C\,{\sc ii}] or CO lines will be a key to directly constrain $f_{\rm gas}$ and confirm the gas mass from the dynamical masses.
For instance, [C\,{\sc ii}] line can be detected in 3--15\,hours source integration in case of $f_{\rm gas}\sim0.9$ based on the calibration of $L_{\rm [CII]}/M_{\rm mol}$ in \citet{2018MNRAS.481.1976Z}\footnote{The calibration in \citet{2018MNRAS.481.1976Z} do not cover the metallicity range of $Z\lesssim0.2\,Z_{\odot}$ \citep[see also,][]{2014A&A...563A..31R}. The estimation of [C\,{\sc ii}] line luminosity here may be optimistic as [C\,{\sc ii}] line is likely to be fainter in low-$Z$ galaxies \citep[e.g.,][]{2019PASJ...71...71H,2020ApJ...896...93H,2024MNRAS.532.2270B}}.

In addition to grain growth, dust destruction can play a critical role, particularly by SN reverse shocks.
Figure \ref{fig:productionmodel} shows the evolution of stellar and dust masses from supernovae (SNe) and asymptotic giant branch (AGB) stars, based on each galaxy's SFH derived using {\sc Bagpipes} or {\sc Prospector} \citep[][see also section \ref{sec:obs}]{2024NatAs.tmp..258Z,2024arXiv240920533C,2024arXiv240618352H}. 
The SFHs are parameterized with an exponential profile (${\rm SFR}(t)\propto\exp(t/\tau)$), with star formation starting at $z=20$ for GHZ2 and $z=30$ for GS-z14-0. 
Our predictions are based on a recipe presented in \citet[][see also, \citet{2024A&ARv..32....2S} for recent review]{2009MNRAS.397.1661V} with the SN dust yield from \citet{2007MNRAS.378..973B} and \citet{2008A&A...479..453Z}, and the AGB dust yield from \citet{2006A&A...447..553F} and the ATON code \citep{2019MNRAS.486.4738D}.
Given the metal-poor nature of the galaxies at $z>12$, we use the dust yield model with the metallicity of $10^{-4}$--$0.1\,Z_{\odot}$.
We integrate the dust yield from stars ranging $0.1$--$100\,M_{\odot}$ under a Larson \citep{1998MNRAS.301..569L} IMF with $\alpha=1.35$ and $M_{\rm ch}=0.35$, which is similar to a Salpeter IMF \citep{1955ApJ...121..161S}.

Given the young stellar ages derived from SED fitting ($\lesssim150\,{\rm Myr}$) and the age of the universe at these redshifts ($\lesssim300\,{\rm Myr}$), dust from SNe dominates the bulk mass.
For GS-z14-0, a high $>80\%$ destruction efficiency by SN reverse shocks is necessary to reproduce the observed dust-to-stellar mass ratios, while destruction is not necessarily significant for GHZ2 (see Figure \ref{fig:productionmodel}).
The surviving dust mass fraction of $<(100-80)=20\%$ in case of GS-z14-0 is actually consistent with 
theoretical studies \citep[e.g.,][]{2007ApJ...666..955N,2016A&A...589A.132B,2019MNRAS.489.4465K}.
The expected significant dust destruction in GS-z14-0 by SN reverse shock contrasts with that inferred for dust-rich galaxies at $z\sim7$ \citep{2019A&A...624L..13L}. 
While inefficient dust destruction or high condensation efficiency may play a more important role in these lower redshift galaxies \citep[][see also, \citealt{2024MNRAS.528.2407P}]{2022MNRAS.512..989D}, it is also possible that the large dust content in REBELS galaxies may be the result of efficient ISM grain growth given their longer ages of tens to hundreds of Myr compared to our $z>12$ systems.

We also found that a lower dust survival rate is necessary to explain the upper limits if the IMF is more top-heavy.
Specifically, $\lesssim4\%$ dust survival rate is required in GS-z14-0 under $M_{\rm ch}=10$ instead of $M_{\rm ch}=0.35$ in the Larson IMF (comparable to the top-heavy IMF), which is smaller than the expectations from several models \citep[e.g.,][]{2015MNRAS.454.4250M}.
Although some theoretical models predict a much lower dust survival rate \citep[e.g.,][]{2007ApJ...666..955N,2020ApJ...902..135S}, it may suggest that the top-heavy IMF scenario is not preferred to explain the overabundance of UV-bright galaxies at $z>10$ from their low dust contents.

\subsection{Star formation efficiency}\label{subsec:SFE}

In this section, we compare efficiencies of gas consumption at $z>12$ with those at $z<9$.
In our fiducial calculations, the total gas masses of the target galaxies are computed as $M_{\rm gas}=M_{\rm dyn}-M_{\ast}$ as outlined in section \ref{subsec:dustprod} and, following \citet{2024A&A...682A..24A}, we assume the ISM is dominated by the molecular gas ($M_{\rm gas}\simeq M_{\rm mol}$). As in section \ref{subsec:dustprod}, we also calculate gas masses assuming $f_{\rm gas}=0.1$--0.9. The caveats associated with these assumptions will be discussed below.
On the other hand, we calculate the SFRs by averaging the non-parametric SFHs over timescales of $10\,{\rm Myr}$ and $100\,{\rm Myr}$ (Table \ref{tab:tab1}).

\begin{figure*}[htbp]
\begin{center}
\includegraphics[width=14cm,bb=0 0 1000 650, trim=0 1 0 0cm]{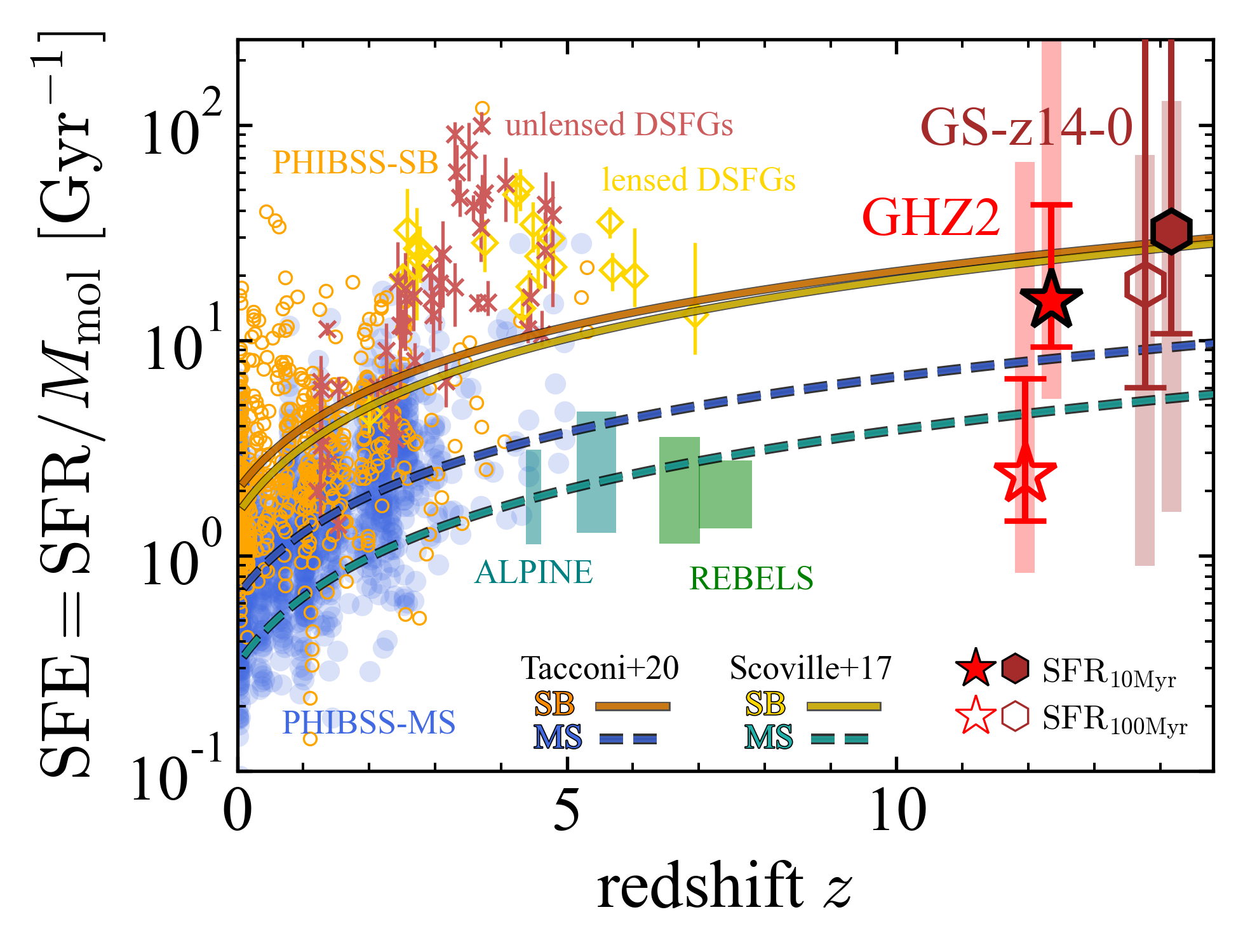}
\caption{Redshift evolution of the star formation efficiency. 
Our fiducial calculations (red stars and brown hexagons) are shown based on ${\rm SFR}_{\rm 10Myr}$ (filled) and ${\rm SFR}_{\rm 100Myr}$ (open markers), respectively. 
We also illustrate SFEs based on the gas masses assuming $f_{\rm gas}=0.1$--0.9 in red and brown shaded regions associated with each marker.
Shifts have been added in the x-axis direction for ease of viewing.
Previous studies of the MS galaxies from PHIBSS \citep[blue circles,][]{2018ApJ...853..179T}, ALPINE \citep[blue shades,][]{2020A&A...643A...5D}, and REBELS \citep[green shades,][]{2024A&A...682A..24A} as well as SB galaxies from PHIBSS (orange circles), unlensed DSFGs \citep[brown crosses,][]{2021MNRAS.501.3926B}, lensed DSFGs \citep[yellow diamonds,][]{2016MNRAS.457.4406A,2021ApJ...921...97J,2022ApJ...933..242Z} are also plotted.
The predicted evolution in \citet[][orange solid and blue dashed lines]{2020ARA&A..58..157T} and \citet[][yellow solid and green dashed lines]{2017ApJ...837..150S}.}
\label{fig:sfe}
\end{center}
\end{figure*}

Figure \ref{fig:sfe} illustrates the evolution of the star formation efficiency (SFE), defined as ${\rm SFR}/M_{\rm mol}$.
For comparison, we plotted data from main-sequence (MS) and `starburst' (SB) galaxies from PHIBSS \citep{2018ApJ...853..179T}, lensed DSFGs \citep{2016MNRAS.457.4406A,2021ApJ...921...97J,2022ApJ...933..242Z}, unlensed DSFGs \citep{2021MNRAS.501.3926B}, ALPINE \citep{2020A&A...643A...5D} and REBELS \citep{2024A&A...682A..24A}.
We also show the scaling relations from \citet{2020ARA&A..58..157T} and \citet{2017ApJ...837..150S} at the stellar mass of $M_{\ast}=10^{10}\,M_{\odot}$. 
While the range of stellar masses at each redshift spreads $\log M_{\ast}/M_{\odot}=8$--11 (with high redshift galaxies showing systematically smaller stellar masses), the dependence of the SFE on the stellar mass is small ($\propto1/10^{0.01-0.03}$).
Also, it is worth noting that molecular gas tracers differ across studies (CO in PHIBSS and SMGs, [C\,{\sc ii}] in ALPINE and REBELS), but these variations do not introduce systematic uncertainties with the dynamical mass-based estimations \citep{2020A&A...643A...5D,2021MNRAS.501.3926B,2024A&A...682A..24A}.

%
%
%
%
%
%

For GHZ2, the SFR averaged over $100\,$Myr,  ${\rm SFR}_{\rm 100\,Myr}$, is smaller than ${\rm SFR}_{\rm 10\,Myr}$, owing to its inferred rapidly increasing star formation activity (see Figure \ref{fig:productionmodel}).
The ${\rm SFR}_{\rm 100\,Myr}$ of GHZ2 is consistent with the expected evolution of MS galaxies, but the most recent star-forming episode, represented by  ${\rm SFR}_{\rm 10\,Myr}$ shows a higher efficiency in better agreement with the so-called ``starbursts'' systems.
In contrast, GS-z14-0 has a comparable high SFE at both 10\,Myr and 100\,Myr timescales (given its more smoothly-rising SFH), indicating a relatively longer-term starburst mode.
The SFEs of GHZ2 and GS-z14-0 in 10\,Myr timescale are ${\rm SFH}_{\rm 10\,Myr}\gtrsim10\,{\rm Gyr}^{-1}$ and are in a good agreement with the expected redshift evolution of starburst galaxies.
This contrasts with the UV-bright galaxies at $z=4$--8  selected via the Lyman break method, which has SFEs of $\sim1\,{\rm Gyr}^{-1}$ and follows the MS scaling relation or even prefers a shallower evolution \citep{2020A&A...643A...5D,2024A&A...682A..24A}.
Indeed, this high SFE in GHZ2 is in line with the interpretation in \citet{2024arXiv241103593Z}, H\,{\sc ii} galaxy-like nature of GHZ2.

The timescales over which the star formation is being calculated can be important when comparing our measurements with those at lower redshifts. 
In Figure \ref{fig:sfe}, the SFRs of UV-selected galaxies at $z=4$--8 (i.e., ALPINE/REBELS galaxies) are calculated from rest-frame UV and IR continuum tracing the average star formation activity within $\sim100\,{\rm Myr}$.
Indeed, \cite{2024MNRAS.533.1111E} reported that  around $10\%$ of $z\approx6$--9  UV-bright galaxies with $M_{\rm UV}\sim-20\,{\rm mag}$ show  an order of magnitude higher SFR in $<10\,{\rm Myr}$. 
Therefore, $\sim10\%$ of the ALPINE/REBELS galaxies may have an order of magnitude larger SFR within a $<10\,{\rm Myr}$ timescale, implying a higher SFR in better agreement with the SB scaling relation.\footnote{A fraction of galaxies having rising SFHs may increase in galaxies with the small absolute UV magnitude \citep{2024MNRAS.533.1111E,2024MNRAS.52711372A,2024arXiv240618352H}. 
As the ALPINE and REBELS galaxies have smaller absolute UV magnitude ($M_{\rm UV}\sim-21$--$-22\,{\rm mag}$) than those at \citet[][$M_{\rm UV}\sim-20\,{\rm mag}$]{2024MNRAS.533.1111E}, the fraction of the galaxies lying on the SB sequence in 10\,Myr timescale may be higher than $\sim10\%$ in the ALPINE and REBELS galaxies.
However, at least not every galaxy has like that rising SFHs \citep{2022MNRAS.516..975T,2024MNRAS.528.2407P}.}
On the other hand, as described above,  both GHZ2 and GS-z14-0 are likely under a starburst episode, which may suggest that this `starburst' phase with efficient gas consumption is more ubiquitous among UV-bright galaxies at $z>12$ than at $z\sim6$.

As described in \citet{2024arXiv240920533C}, GS-z14-0 exhibits a smaller gas fraction ($M_{\rm gas}/(M_{\ast}+M_{\rm gas})\sim0.3$) than those at $z=4$--8 \citep[$M_{\rm gas}/M_{\ast}\sim1$][]{2020A&A...643A...5D,2024A&A...682A..24A}, indicating small gas reservoir compared with an existing stellar mass.
Approximating $M_{\rm gas}\approx f_{b}M_{\rm halo}$, where $f_b$ and $M_{\rm halo}$ is the cosmic baryon fraction and the host halo mass, respectively, the stellar-to-halo mass ratio ($\epsilon_{\ast}=M_{\ast}/M_{\rm halo}$) becomes around 0.5, which is consistent with recent studies implying high $\epsilon_{\ast}$ \citep[e.g.,][]{2024ApJ...965...98C,2024arXiv241008290S}.
The high SFE and small gas fraction in GS-z14-0 align also with the feedback-free star formation model \citep{2023MNRAS.523.3201D,2024A&A...690A.108L}. All these scenarios might be (at least partially) responsible for the reported overabundance of UV-bright galaxies at $z\gtrsim10$.
We thus hypothesize that GS-z14-0 is converting most of its molecular gas into stars in a recent `starburst', and will run out the molecular gas \citep[or expel the gas and dust components, see][]{2024A&A...684A.207F}, and turn off their star formation within the next several $10\,{\rm Myrs}$.

On the other hand, GHZ2 shows a large fraction of molecular gas mass within the baryonic content ($M_{\rm gas}/(M_{\ast}+M_{\rm gas})\sim0.7$). Its instantaneous high SFE and high gas fraction may be the result of a bursty star formation history, which increases the visibility of UV-bright galaxies during short and intense bursts of star formation \citep[e.g.,][]{2023ApJ...955L..35S,2024arXiv240504578K}.
Further observations capturing rest-frame optical stellar continuum will enable us to make more accurate constraints on the stellar mass, and therefore on the gas fraction.


One major caveat in using $M_{\rm dyn}-M_{\ast}$ as a proxy of $M_{\rm gas}$ is the uncertain fraction of the molecular gas within the total gas.
However, if the molecular gas fraction is below unity, the derived $M_{\rm mol}$ decreases, and the SFE increases.
Therefore, the adopted molecular gas fraction does not affect our conclusion.
We acknowledge, however, that conducting direct observations of cold ISM tracers, such as [C\,{\sc ii}] or CO lines, is crucial to put better constraints on the molecular gas mass.
As shown in the shaded boxes in Figure \ref{fig:sfe}, the assumption of the $f_{\rm gas}$ range of 0.1--0.9 prevents imposing any meaningful constraints on the SFE.
The detection experiment of [C\,{\sc ii}] will be an important key to confirm or rule out our conclusion.

%
%
%
%
%
%
%
%
%
%
%

%
%
%
%
%
%
\section{Summary}\label{sec:summary}
In this paper, we have investigated the dust and gas content of the brightest galaxies at $z>12$, GHZ2, and GS-z14-0, based on direct ALMA observations that trace the bulk of the dust mass in these galaxies.
From the combination of ultra-deep ALMA observations and an analytical thermal equilibrium model, we have found:

\begin{itemize}
\item The $3\sigma$ upper limits on the dust continuum at $\lambda_{\rm rest}=88\,\mu{\rm m}$ are $10.3\,\mu{\rm Jy}$ and $14.9\,\mu{\rm Jy}$ for Band-6 observations on GHZ2 and GS-z14-0, respectively. 
Combining these observations with an analytical model based on thermal equilibrium and clumpy dust geometry \citep{2020MNRAS.495.1577I}, we derived dust mass upper limits of $\log M_{\rm dust}/M_{\odot}<5.0$ and $<5.3$, respectively. 

\item The comparison of our constraints on $M_{\ast}$--$M_{\rm dust}$ plane and dust production models at $z>4$ do not imply efficient dust grain growth with extreme condensation efficiency $f_{\rm cond}$. 
Assuming most massive galaxies at $z>12$ are potential progenitors of most massive galaxies at $z=7$, a short accretion timescale scenario is supported rather than a high $f_{\rm cond}$ scenario in \citet{2019MNRAS.490..540L}.
Initial dust size distributions with a moderate amount of large-size grains, resulting in $\tau_{\rm acc}\sim10\,{\rm Myr}$, are preferred in our constraints on the $Z$--D/G relation. 

\item Given the very young ages of the target galaxies, dust yield from AGB stars and grain growth are not likely to be effective. The observed dust-to-stellar mass ratios in GS-z14-0 can be naturally explained by the dust survival rate of $<20\%$ from the reverse shocks by SN. 

\item Based on the gas masses from $M_{\rm gas}=M_{\rm dyn}-M_{\ast}$ and SFHs, we inferred star formation efficiencies within the past 10\,Myr, which reach high values above $\sim10\,{\rm Gyr}^{-1}$, comparable with the predicted SFEs' evolution of starburst galaxies. GHZ2 may exhibit short starburst-like episodes only for relatively short periods of up to $\sim10\,{\rm Myr}$, while GS-z14-0 is likely to have such starburst mode in a relatively long term of $\sim100\,{\rm Myr}$. From a comparison with UV-bright galaxies at $z\sim6$, we hypothesize that such  `starburst' activity may be more ubiquitous in UV-bright galaxies at $z>12$ than at $z\sim6$, potentially due to bursty SFHs \citep[e.g.,][]{2023ApJ...955L..35S} and/or negligible feedback coming from super-Eddington star formation \citep[e.g.,][]{2023MNRAS.523.3201D}. 

\end{itemize}

Our study demonstrates the importance of observing cold interstellar medium tracers, such as dust continuum and [C\,{\sc ii}] line observations with ALMA to derive the current and future star formation activity of UV-bright galaxies.
Since our analysis with the thermal equilibrium model predicts very high $T_{\rm dust}$, further deep ALMA observations at high frequency to capture near the peak of dust thermal emission will be useful to identify dust at $z>12$.
For instance, $\sim30\,{\rm hours}$ ALMA Band 9 observations covering rest-frame $\sim30\,\mu{\rm m}$ allows us to add tighter constraint on $M_{\rm dust}$ at $T_{\rm dust}>100\,{\rm K}$ regime.
To further understand the emergence of dust across the Universe, investigating the gas and dust content of most massive galaxies at $z\sim8$--11 is necessary to constrain the grain growth in the ISM.
Systematic comparison of ISM conditions, such as gas/electron density, across the Universe, might be critical to understanding the origin of the lack of dust and the potential `starburst' nature of $z>12$ galaxies.

\acknowledgments
We thank P. Dayal for providing us with data from their semi-analytical calculation. This paper makes use of the following ALMA data: ADS/JAO.ALMA\#2021.A.00020.S, \#2023.A.00017, \#2023.A.00037.
ALMA is a partnership of ESO (representing its member states), NSF (USA), and NINS (Japan), together with NRC (Canada), MOST and ASIAA (Taiwan), and KASI (Republic of Korea), in cooperation with the Republic of Chile. The Joint ALMA Observatory is operated by ESO, AUI/NRAO, and NAOJ.
IM and AKI acknowledge support from NAOJ ALMA Scientific Research Grant number 2020-16B.
The authors acknowledge the GLASS and JADES teams led by Tommaso Treu and Daniel Eisenstein \& Nora Luetzgendorf, respectively, for developing their observing programs. 
The {\it JWST} data presented in this article were obtained from the Mikulski Archive for Space Telescopes (MAST) at the Space Telescope Science Institute. 
The specific observations analyzed can be found in the MAST \dataset[10.17909/wz7w-2208]{http://dx.doi.org/10.17909/wz7w-2208}.
Data analysis was in part carried out using the Multi-wavelength Data Analysis System operated by the Astronomy Data Center (ADC), National Astronomical Observatory of Japan.
KK acknowledges support from the Knut and Alice Wallenberg Foundation (KAW 2020.0081).
PS acknowledges INAF Mini Grant 2022 "The evolution of passive galaxies through cosmic time".
\renewcommand{\thesection}{A}


%
%
%
%
%
%
%
%
%
%
%
%

\clearpage
\bibliography{main.bib}{}
\bibliographystyle{apj.bst}

\end{document}